\documentclass[aps,pra,twocolumn,superscriptaddress,showpacs,floatfix]{revtex4-1}

\usepackage{times,amssymb,amsmath}
\usepackage{array,color}
\usepackage{ifpdf}
\usepackage{tensor}
\usepackage[varg]{txfonts} 
\usepackage[pdftex]{graphicx}
\renewcommand{\vec}[1]{\mathbf{#1}}

\mathchardef\mhyphen="2A

\newcommand{\tns}[2]{\tensor[^#1]{#2}{}}
\begin{document}


\title{Accessing electronic correlations by half-cycle pulses and time-resolved spectroscopy}
\author{Y. Pavlyukh}
\email[]{yaroslav.pavlyukh@physik.uni-halle.de}
\author{J. Berakdar}
\affiliation{Institut f\"{u}r Physik, Martin-Luther-Universit\"{a}t
  Halle-Wittenberg, 06120 Halle, Germany}
\date{\today}
\begin{abstract}
%
Ultrashort non-resonant electromagnetic pulses applied to effective one-electron systems
may operate on the electronic state as a position or momentum translation operator. As
derived here, extension to many-body correlated systems exposes qualitatively new aspects.
For instance, to the lowest order in the electric field intensity the action of the pulse
is expressible in terms of the two-body reduced density matrix enabling thus to probe
various facets of electronic correlations.  As an experimental realization we propose a
pump-probe scheme in which after a weak, swift "kick" by the non-resonant pulse the
survival probability for remaining in the initial state is measured.  This probability we
correlate to the two-body reduced density matrix. Since the strength of electronic
correlation is bond-length sensitive, measuring the survival probability may allow for a
direct insight into the bond-dependent two-body correlation in the ground state. As an
illustration, full numerical calculations for two molecular systems are provided and
different measures of electronic correlations are analyzed.
%
%
\end{abstract}
\maketitle
\section{Introduction}
Recent experiments demonstrated the feasibility of ultrashort attosecond laser pulses and
their use in combination with other optical pump-probe techniques to access various facets
of the electronic and ionic
dynamics~\cite{goulielmakis_single-cycle_2008,ding_generation_2009,bergues_attosecond_2012,
  kampfrath_resonant_2013,chini_generation_2014}.  Of a special relevance here are
strongly time-asymmetric pulses. The simplest example is a mono cycle, time asymmetric
pulse for which one half cycle could be short and strong and the other half cycle could be
longer but weaker in a way that the time integral over the electric field amplitude
vanishes, as required for a propagating pulse.  When interacting with matter, the effect
of two half cycles could be quite different
~\cite{jones_ionization_1993,raman_ionization_1996,ahn_quantum_2001,wu_giant_2012}.  For
instance, if the first half cycle (with duration $\tau$) is shorter than the time of
relevant transitions taking place in the system, whilst the second, weaker half cycle is
much longer, then the effect of the whole pulse is mainly governed by the first half cycle
of the pulse (HCP). Henceforth we refer to this situation as the HCP case.  The second
half cycle may act as an off-set DC tail (if weak/long enough).
%
 In this limit it is convenient and sometime sufficient to introduce theoretically the
 notion of HCP kick as a useful idealization of the action of the whole
 pulse~\cite{jones_ionization_1993,reinhold_ionization_1993,jones_creating_1996,
   matos-abiague_sustainable_2003,matos-abiague_controlling_2003,stapelfeldt_colloquium:_2003,
   matos-abiague_ultrafast_2004,matos-abiague_ultrafast_2005,matos-abiague_femtosecond_2004,
   matos-abiague_emission_2005,moskalenko_revivals_2006,mestayer_transporting_2007,zhu_photoinduced_2008}:
 Namely, in the HCP case, the system is insensitive to the details of the pulse temporal
 shape. This allows relating the action of HCP to that of a kick, i.e., a $\delta$
 function in time.  Substantial simplifications of the triggered quantum dynamics follow
 then.  We will demonstrate below that by performing measurements in an appropriate setup
 on the system immediately after such an excitation allows to infer the strength of the
 electronic correlation in the system and its dependency on the internal structure, and to
 quantify the entanglement. The experimental arrangement that we suggest is the following:
 We start from the ground state, of say some molecular structure (another stationary case
 is also possible) and "shake" the electronic system by a weak but short HCP.  As the
 system evolves and possibly dissociates we measure, e.g., via the time-resolved
 photoemission~\cite{stolow_femtosecond_2004} the initial-state remaining, i.e. what we
 will call the \emph{survival probability}.  Detecting only the kinetic energies of the
 dissociating ions upon the kick, one may access information on the ionic time evolution
 that can be used when measuring the time-resolved survival probability as a function of
 the molecular bond (since at this particular time one has information on the molecular
 bond distance).  Currently we are not aware of any corresponding experiment to this
 proposal, however, the ingredients of the suggested setup, such as HCP, time-resolved
 photoemission were demonstrated
 ~\cite{sansone_electron_2010,hockett_time-resolved_2011,schnorr_time-resolved_2013}.
 From the survival probability we deduce features akin to the reduced two-particle density
 matrix.

Specifically, the envisaged experiment should measure upon the kick the initial
state occupation probability as a function of the evolving internuclear distance.  It is
this quantity which we calculated numerically and will be discussing below.  Knowing (from
theory or experiment~\cite{jiang_ultrafast_2013,ibrahim_tabletop_2014}
molecular bond distance upon the non-resonant pulse excitation we can so image the time
evolution of the survival probability. We relate it below to the reduced two-particle
density matrix which vanishes for the single determinantal states and, thus, can be used
as a measure of electronic correlations.

So the key point of this study is how to measure and quantify electronic correlations. In
Sec.~\ref{sec:theory} we give an explicit expression for the probability of the system to
remain in its ground state after the application of a $\delta$-like pulse. While for
one-electron systems the quantum dynamics was discussed by many authors the nontrivial
part tackled here is the many-body nature of the problem. In Sec.~\ref{sec:measure} we put
the discussed probability in the context of other proposed measures of electronic
correlations: the Frobenius norm of the second \emph{cumulant matrix} and the \emph{von
  Neumann entropy}.  Stretching molecular bonds incur in general a change in the amount of
correlations in the electronic subsystem. The proposed experiment is illustrated by
numerical simulations which show what a possible outcome may be expected
(Sec.~\ref{sec:study}). The dependence of different correlation measures on the
interatomic distances is studied, and analytical expressions in the asymptotic regime are
obtained. Additionally, our calculations allow to test numerically for important
inequalities for the entropy measures.
\section{Theory \label{sec:theory}}
Consider an $N$-electron system in its ground state. Its electronic properties are
completely described by the many-body wave function $\psi_0(1,2,\ldots,N)$. Although the
choice of the gauge is irrelevant for the present discussion we will assume here the
light-matter interaction to be given in the length gauge by the dipole operators
$\hat{d}^\alpha$, where $\alpha$ denotes a projection (determined by the pulse polarization).
The coupling operators are of
one-particle type, Hermitian ($\hat{d}^\alpha=(\hat{d}^\alpha)^\dagger$) and of vector
character. By choosing a suitable one-particle basis we can cast the light-matter
interaction in the second-quantized form $\delta H(t)=-\sum_\alpha \sum_{i,j}d_{ij}^\alpha
E^\alpha(t)\hat c_i^\dagger \hat c_j$, where $E^\alpha$ are the components of the electric
field vector. Acting on the system with a pulse that has the time-dependence
$E^\alpha(t)=E_0^\alpha \delta_\epsilon(t)$ with a peak amplitude at $t=0$,  we can
write the state of the system at $t=\epsilon$ shortly after the pulse as
\begin{equation}
\label{eq:exp}
\psi_+=e^{i\sum_{i,j}S_{ij} \hat c_i^\dagger \hat c_j}\psi_0=e^{i\hat S}\psi_0.
\end{equation}
$\delta_\epsilon$ is a regularized version of the mathematical $\delta$-function on
$[0,\epsilon]$ interval and $S_{ij}=-\int_0^\epsilon dt \sum_\alpha d_{ij}^\alpha
E_0^\alpha \delta_\epsilon(t)$ has a dimension of the action (we use atomic units
throughout the text). Mapping the pulse generated state on the ground (initial) state
$\psi_0$, i.e., taking the overlap $\langle \psi_0|\psi_+\rangle$ of \eqref{eq:exp} we
obtain the survival (recurrence) probability to be in $\psi_0$ after the pulse. This
quantity is central to the following discussion.

 The \emph{quantum-mechanical} average of the operator $\hat S=\sum_{i,j}S_{ij} \hat
 c_i^\dagger \hat c_j$ can be treated like any statistical average using the
 \emph{cumulant expansion}:
\begin{equation}
\label{eq:cumulant}
\langle \psi_0|\psi_+\rangle=\exp\Bigl\{\frac{i}{1!}\mathfrak{S}_1
+\frac{i^2}{2!}\mathfrak{S}_2+\frac{i^3}{3!}\mathfrak{S}_3+\cdots\Bigr\}.
\end{equation}
The correlation functions are well known:
\begin{subequations}
\label{eq:corr}
\begin{eqnarray}
\mathfrak{S}_1&=&\langle \hat S\rangle,\quad
\mathfrak{S}_2=\langle \hat S^2\rangle-\langle \hat S\rangle^2,\\
\mathfrak{S}_3&=&
\langle \hat S^3\rangle-3\langle \hat S^2\rangle\langle \hat S\rangle+2\langle \hat S\rangle^3.
\end{eqnarray}
\end{subequations}
These averages can be computed in terms of reduced density matrices (RDM)
\begin{subequations}
\label{eq:defD}
\begin{eqnarray}
{^1\!D^i_j}&=&\langle \psi_0|\hat c_i^\dagger \hat c_j|\psi_0\rangle,\\
{^2\!D^{ik}_{jl}}&=&\langle \psi_0|\hat c_i^\dagger \hat c_k^\dagger \hat c_l\hat c_j|\psi_0\rangle,\\
{^3\!D^{ikm}_{jln}}&=&\langle \psi_0|\hat c_i^\dagger \hat c_k^\dagger \hat c_m^\dagger \hat c_n\hat c_l\hat c_j|\psi_0\rangle.
\end{eqnarray}
\end{subequations}
In order to do so we need some additional notations for operators. $\hat{S}^n$ is a
$n$-body operator, i.e. it is given by an expression containing $n$ creation and $n$
annihilation operators. Let $[\hat{S}^n]$ denote a one-particle operator which is given by
the $n$th power of  $\hat{S}$ operator in the first quantization, i.e.
\[
[\hat{S}^n]=\sum_{ij}(S^n)_{ij} \hat c_i^\dagger \hat c_j.
\]
In these notations:
\begin{eqnarray*}
\mathfrak{S}_2&=&-\langle\hat S\otimes\hat S\rangle+\langle [\hat S^2]\rangle
-\langle[\hat S]\rangle^2,\\
\mathfrak{S}_3&=&-\langle \hat S\otimes \hat S\otimes\hat S\rangle-3\langle[\hat S^2]\otimes \hat S\rangle
+\langle[\hat S^3]\rangle\\
&&\quad+3\langle \hat S\otimes\hat S\rangle\langle\hat S\rangle
-3\langle [\hat S^2]\rangle\langle[\hat S]\rangle+5\langle[\hat S]\rangle^3,
\end{eqnarray*}
where the ${}^p\!\hat{A}\otimes{}^q\!\hat{B}$ denotes the \emph{normal form} of the product
$^p\!\hat{A}$ and $^q\!\hat{B}$, the $p$- and $q$-body operators, respectively.  It is
defined as follows:
\begin{eqnarray*}
^p\!\hat{A}&=&\sum_{\vec{i},\vec{j}}A^\vec{i}_\vec{j}\hat c^\dagger_{i_1}\ldots \hat c^\dagger_{i_p}\left(\hat c_{j_1}\ldots \hat c_{j_p}\right)^T,\\
^q\!\hat{B}&=&\sum_{\vec{k},\vec{l}}B^\vec{k}_\vec{l}\hat c^\dagger_{k_1}\ldots \hat c^\dagger_{k_q}\left(\hat c_{l_1}\ldots \hat c_{l_q}\right)^T,\\
{}^p\!\hat{A}\otimes{}^q\!\hat{B}&=&\sum_{\vec{i},\vec{j}}\sum_{\vec{k},\vec{l}}A^\vec{i}_\vec{j}B^\vec{k}_\vec{l}
\hat c^\dagger_{i_1}\ldots \hat c^\dagger_{i_p}\hat c^\dagger_{k_1}\ldots \hat c^\dagger_{k_q}\\
&&\hspace{7em}\times\left(\hat c_{l_1}\ldots \hat c_{l_q}\hat c_{j_1}\ldots \hat c_{j_p}\right)^T.
\end{eqnarray*}
In these expressions $\vec{i}$, $\vec{j}$, and $\vec{k}$, $\vec{l}$, are the $p$ and
$q$-dimensional vectors of indices.

For our discussion it is instructive to introduce the \emph{cumulant density
  matrices}~\cite{mazziotti_reduced-density-matrix_2007,mazziotti_two-electron_2012}
${^p\!\Delta}$ which allow to decompose the $p$-RDM in terms of correlated (connected)
$p$-particle correlator and products of lower order
correlators~\cite{skolnik_cumulant_2013}:
\begin{subequations}
\label{eq:defCRD}
\begin{eqnarray}
\frac{1}{1!}{^1\!D^i_j}&=&{^1\!\Delta^i_j},\\
\frac{1}{2!}{^2\!D^{ik}_{jl}}&=&{^1\!\Delta^i_j}\wedge{^1\!\Delta^k_l}+{^2\!\Delta^{ik}_{jl}},\\
\frac{1}{3!}{^3\!D^{ikm}_{jln}}&=&{^1\!D^i_j}\wedge{^1\!D^k_l}\wedge{^1\!D^m_n}
+3\,{^2\!\Delta^{ik}_{jl}}\wedge{^1\!D^m_n}+{^3\!\Delta^{ikm}_{jln}},
\end{eqnarray}
\end{subequations}
where $\wedge$ denotes the \emph{wedge product}~\cite{mazziotti_approximate_1998} (for
mathematical details  we refer to a treatise on differential
forms~\cite{flanders_differential_1989} where  $\wedge$ appears under the name exterior product).
We have, for instance:
\[
{^1\!\Delta^i_j}\wedge{^1\!\Delta^k_l}=\frac12\left({^1\!\Delta^i_j}{^1\!\Delta^k_l}
-{^1\!\Delta^i_l}{^1\!\Delta^k_j}\right).
\]
The $1/n!$ prefactor on the rhs of Eqs.~\eqref{eq:defCRD} appears naturally when the
density matrices are written in the first quantization:
\begin{multline}
\frac{1}{n!}{^n\!D(1,\ldots,n;1^\prime,\ldots,n^\prime)}=\int\psi_0^*(1,\ldots,n,n+1,\ldots,N)\\
\times\psi_0(1^\prime,\ldots,n^\prime,n+1,\ldots,N) d(n+1,\ldots,N),
\end{multline}
where $i\equiv(r_i,s_i)$ denotes a collection of space and spin coordinates.

Returning back to Eq.~\eqref{eq:cumulant} we find by direct comparison
$\frac{1}{1!}\mathfrak S_1=\sum_{ij}S_{ij}{^1\!D^i_j}$. It is, however, the second
cumulant that gives the lowest order (in $|\vec E_0|^2$) contribution to the survival
probability, namely
\begin{eqnarray}
|\langle \psi_0|\psi_+\rangle|^2&\approx&\exp{[-\mathfrak S_2]}\approx
1-\mathfrak S_2,\nonumber\\
\mathfrak S_2&=&\sum_{ijkl}S_{ij}S_{kl}{^2\!D^{ik}_{jl}}+\sigma^2_S.
\label{eq:res1}
\end{eqnarray}
Thus, $\mathfrak S_2$ can be written as the averaged value of a two-body operator (first
term in Eq.~\eqref{eq:res1}), whereas $\sigma^2_S=\langle [\hat S^2]\rangle -\langle[\hat
  S]\rangle^2$ is computed from 1-RDM. This equation can be written in an alternative
form. Consider the natural orbital basis (i.e. a basis in which 1-RDM is diagonal):
\begin{eqnarray}
\mathfrak S_2&=&\sum_{ijkl}S_{ij}S_{kl}\left[{}^2\!\Delta^{ik}_{jl}+f_if_k
(\delta_{ij}\delta_{kl}-\delta_{il}\delta_{kj})\right]\nonumber\\
&+&\sum_{im}(S_{im}S_{mi}f_i-S_{ii}S_{mm}f_if_m)\nonumber\\
&=&\sum_{ijkl}S_{ij}S_{kl}{^2\!\Delta^{ik}_{jl}}+\sum_{im}|S_{im}|^2f_i(1-f_m),\label{eq:res2}
\end{eqnarray}
where $f_i$ is the occupation number of the $i$th natural orbital. Despite the fact that
the second term in Eq.~\eqref{eq:res2} has a form of the \emph{Fermi golden
  rule}~\footnote{There are actually two rules under this name. We are talking here about
  the expression for the total probability which is obtained from the rate equation by the
  integration with the frequency profile of the excitation pulse.} the latter is only
valid for sufficiently long pulses (adiabatic switching).

Eq.~\eqref{eq:res1} is a quite remarkable result as it allows to express the averaged
2-RDM in terms of experimentally measurable quantities: the survival probability and the
mean square deviation of an excitation operator. Alternatively, one can use the
form~\eqref{eq:res2} to access the $^2\!\Delta$ cumulant as a correction to the
single-particle result. A possible experiment that we sketched in the introduction, could
be a pump-probe setup in which the system is excited non-resonantly by a half-cycle pump
pulse and the probe laser pulse is used to monitor the ground state occupation.  We
envisage an application to molecular systems in which the amount of electronic
correlations and, thus, $\langle[\hat S\otimes\hat S]\rangle$ are driven by changes of
geometry. On the other hand, the one-body part $\sigma^2_S$ is a quantity that is weakly
dependent on the geometric configuration (see also the discussion below) and can also be
easily computed~\footnote{We note by passing that the quantities which are defined here
  are of a general nature. In fact, in solids a drastic change in electronic properties
  may occur upon driving the lattice by an external field, as confirmed by series of
  recent experiments (albeit not dedicated to the quantity we are discussing here, the
  survival probability)~\cite{porer_non-thermal_2014}).}.

For a faithful correlation measure it is desirable to minimize the artifacts coming from
the dependence of $\hat{S}$ on the system's geometry. Consider a system subject to
stretching. The values of the matrix elements $S_{ij}$ can be quite large as compared to
the equilibrium geometry. Correspondingly, the value of $\mathfrak S_2$ obtained according
to Eqs.~\eqref{eq:res1},~\eqref{eq:res2} will be at variance with the equilibrium value --
the effect that is not necessarily reflecting properties of $^2\!\Delta$. It is possible,
however, to suppress to some extent the large contributions to $\mathfrak S_2$ originating
from the diagonal matrix elements of $\hat S$. This can be achieved, for example, by
performing the experiment with oriented systems and applying the HCP field in the
direction for which $\langle S\rangle=0$. Such condition is always possible to achieve for
systems which contain $C_s$ as their symmetry subgroup and is quite common. Under this
condition also $\sigma^2_S$ is weakly dependent on the geometry.

Is it possible to devise a measurement that exclusively probes the cumulant density
matrix $^2\Delta$? Our answer to this question is negative, based on the fact that the
generating functional for the cumulant and the reduced density matrices
\begin{equation}
G(J)=\langle\psi_0|O\left(e^{\sum_k{J_k\hat{c}_k^\dagger+J^*_k\hat{c}_k}}\right)|\psi_0\rangle
\label{eq:grassmann}
\end{equation}
is different from the bosonic-like generator of the evolution operator~\eqref{eq:exp}.  In
other words, in order to probe the fermionic RDMs one needs a direct coupling to fermionic
degrees of freedom as Grassmann variables $J_k$ in Eq.~\eqref{eq:grassmann} realize.

Nonetheless, we will demonstrate below using two numerical examples that the survival
probability is a versatile measure of electronic correlations and will compare it to other
proposed measures of electronic correlations and entanglement.
\section{measures of electronic correlation and entanglement\label{sec:measure}}
It was shown by Juh{\'a}sz and Mazziotti~\cite{juhasz_cumulant_2006} that the Frobenius
norm of the second cumulant matrix ($||^2\!\Delta||_F^2=\mathrm{Tr}[(^2\!\Delta)^2]$)
possesses a number of properties that make it a useful measure. $||^2\!\Delta||_F$ scales
linearly with the system size and vanishes for single-determinant states. Although it is
well suited to compare different configurations of the same system such a measure is less
suited  to compare different systems as its upper limit is not known.\\
 Here come
informational measures into play, e.~g. \emph{von Neumann entropy} .  The entropies based
on the 1-RDM have been widely studied. Less known are the entropies based on 2-RDM.
Carlen and Lieb~\cite{carlen_entropy_2014} considered recently the bipartite fermionic
states and proved several bounds for the entropy based on ${^2\!D}$. For a Hilbert space
of the $\mathcal H_1\otimes\mathcal H_2$ Hamiltonian and corresponding $\rho_{12}$ density
matrix the von Neumann entropy can be computed as
\[
S_{12}=-\mathrm{Tr}\rho_{12}\log{\rho_{12}},
\]
where the trace is understood in the sense of the tensor product $\mathcal
H_1\otimes\mathcal H_2$.  The 2-RDM of a $N$-particle fermionic system can be considered
as a density matrix of a bipartite fermionic state. For this case the following bounds are
known (cf. Carlen and Lieb~\cite{carlen_entropy_2014})
\begin{eqnarray}
S(\rho_{12})&\ge& 2\ln{N}+\mathcal{O}(1),\label{eq:ineq1}\\
2S_1-S_{12}&\ge& \ln\left(\frac{2}{1-\mathrm{Tr}\rho_1^2}\right)
\ge\ln\left(\frac{2}{1-e^{-S_1}}\right).\label{eq:ineq2}
\end{eqnarray}
In the following we compare different correlation measures and numerically verify the
inequalities.

\subsection{Applications}

All results so far were represented in some abstract one-particle basis. In what follows
we will focus on molecular systems with equal number of spin up ($N_\alpha$) and spin
down ($N_\beta$) electrons. To treat such systems it is convenient to work in the closed
shell Hartree-Fock (RHF) molecular orbital (MO) basis, distinguish spin up and spin down
MO states and to denote them as $i$ and $\bar{i}$, respectively. This implies some
additional symmetries. Obviously, $^1\!D^{i}_{j}=\,^1\!D^{\bar{i}}_{\bar{j}}$ and
$^1\!D^{i}_{\bar{j}}=\,^1\!D^{\bar{i}}_{j}=0$, and we have the following blocks for the
2-RDMs:
\begin{equation}
^2\!D^{ij}_{kl}=\,^2\!D^{\bar{i}\bar{j}}_{\bar{k}\bar{l}},\quad
^2\!D^{i\bar{j}}_{k\bar{l}}=\,^2\!D^{\bar{i}j}_{\bar{k}l},
\label{eq:notation}
\end{equation}
and same holds for the cumulants. Since 2-RDMs in Eq.~\eqref{eq:notation} are obtained as
scalar products of the $c_ic_j|\psi\rangle$ and $c_ic_{\bar{j}}|\psi\rangle$ vectors we
will denote these blocks as AA and AB. Other 1- and 2-RDMs are obtained by the use of
anti-commutation relations. In order to emphasize spin degrees of freedom we separately
compute the electronic correlation measures associated with AA and AB blocks.

For the computation of entropies the density matrices must be normalized to have trace
equal to one. Thus, we introduce
\begin{eqnarray}
\tns{1}{\tilde{D}}^{i}_{j}&=&\frac{1}{N_\alpha}\tns{1}{D}^{i}_{j},\\
^2\!\tilde{D}^{ij}_{ij}&=&\frac{1}{N_\alpha(N_\alpha-1)}\,^2\!D^{ij}_{ij},\quad
^2\!\tilde{D}^{i\bar{j}}_{i\bar{j}}=\frac{1}{N_\alpha N_\beta}\,^2\!D^{i\bar{j}}_{i\bar{j}},
\end{eqnarray}
with $\mathrm{Tr}[\tns{1}{\tilde{D}}]=\sum_i\tns{1}{\tilde{D}}^{i}_{j}=1$,
$\mathrm{Tr}[\tns{2}{\tilde{D}}]=\sum_{ij}\tns{2}{\tilde{D}}^{ij}_{ij}=1$, and
$\mathrm{Tr}[\,\tns{2}{\tilde{D}}]=\sum_{ij}\tns{2}{\tilde{D}}^{i\bar{j}}_{i\bar{j}}=1$.

Carlen and Lieb showed~\cite{carlen_entropy_2014} that the inequalities
(\ref{eq:ineq1},\ref{eq:ineq2}) are saturated for single Slater determinant states. In
fact, it can be verified that boundary values are achieved for each spin channel
separately. For single determinant states the cumulant matrix $^2\Delta$ vanishes and
2-RDM is given by its unconnected components,
${^2\!D^{ik}_{jl}}={^1\!\Delta^i_j}\wedge{^1\!\Delta^k_l}$. Specifically,
${^2\!D^{ik}_{jl}}=f_if_k(\delta_{ij}\delta_{kl}-\delta_{il}\delta_{jk})$, and
${^2\!D^{i\bar{k}}_{j\bar{l}}}=f_if_{\bar{k}}\delta_{ij}\delta_{\bar{k}\bar{l}}$ in the AA
and AB channels. The first matrix has $N_\alpha(N_\alpha-1)/2$ nonzero eigenvalues equal
to 2, whereas the second one has $N_\alpha N_\beta$ nonzero eigenvalues equal to
1. Correspondingly, the entropies in each channel are $S_{AA}^0=\log{\frac12
  N_\alpha(N_\alpha-1)}=2\log{N_\alpha}+\mathcal{O}(1)$, and $S_{AB}^0=\log{N_\alpha
  N_\beta}$. In the next section we will plot the entropies with respect to these
reference values.
\section{numerical illustrations\label{sec:study}}
\begin{figure}[b!]
\includegraphics[width=0.82\columnwidth]{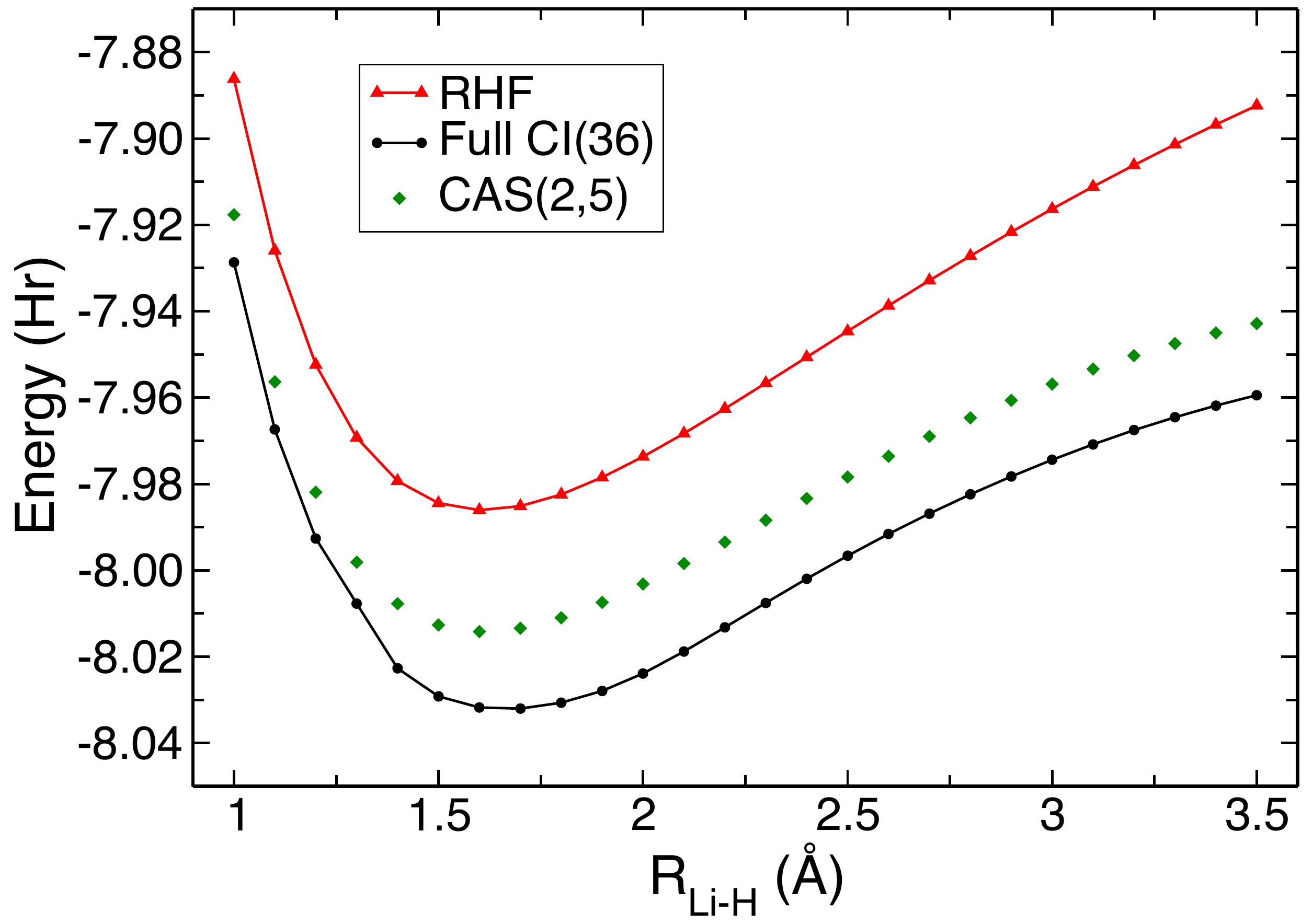}
\caption{(Color online) Potential energy surfaces for the stretched LiH molecule.
\label{fig:LiH_PES}}
\end{figure}
\begin{figure}[t!]
\includegraphics[width=\columnwidth]{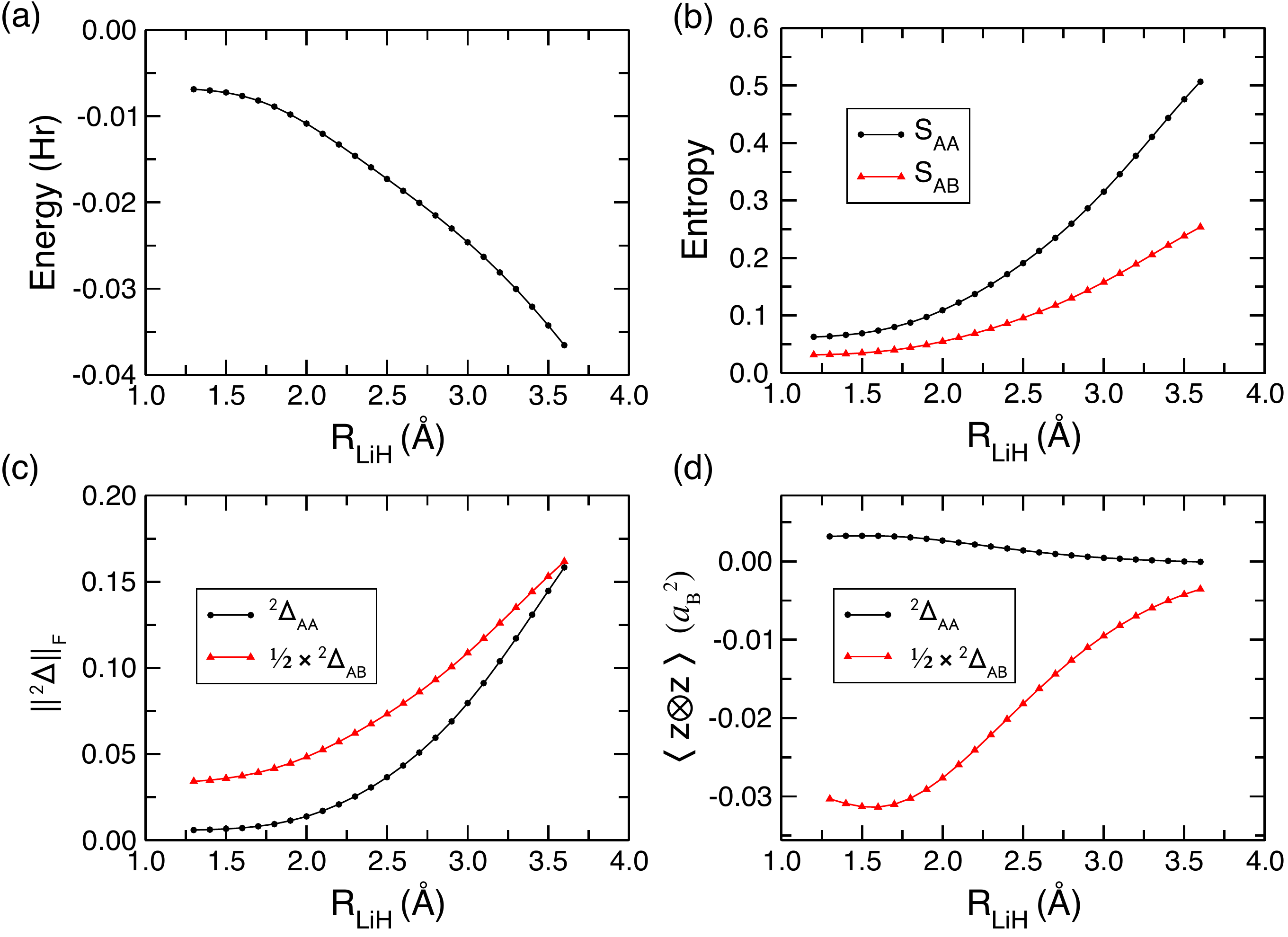}
\caption{(Color online) Full CI calculation for the stretched LiH molecule performed with
  the 6-311++G(2d,2p) basis set, 20 molecular orbitals are included in a correlated
  treatment. The electron correlation increases monotonically as the bond is
  stretched. This is manifested in the correlational energy (a), the entropies (b), the
  Frobenius norm of the second cumulant matrices (c), and the expectation value of the
  $z\otimes z$ operator (cf. $\langle[\hat S\otimes\hat S]\rangle$ in Eq.~\eqref{eq:res1}
  or $\sum_{ijkl}S_{ij}S_{kl}{}^2\!\Delta^{ik}_{jl}$ in Eq.~\eqref{eq:res2}) (d).
    The entropies are given with respect to the single determinant values $S_{AA}^0$ and
    $S_{AB}^0$.
\label{fig:LiH}}
\end{figure}
\begin{figure}[]
\includegraphics[width=0.82\columnwidth]{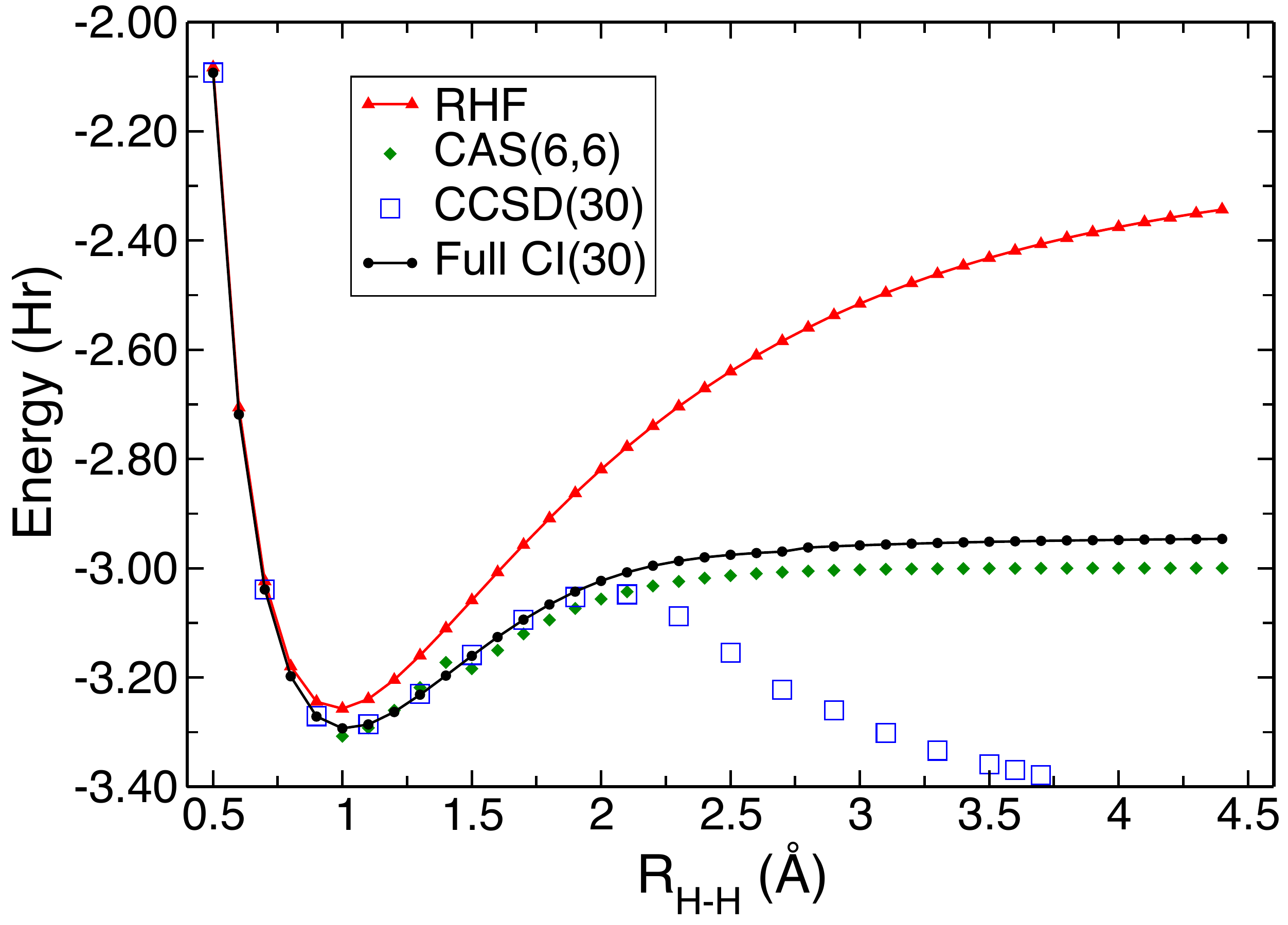}
\caption{(Color online) Potential energy surfaces for the symmetrically stretched H$_6$ molecule.
\label{fig:H6_PES}}
\end{figure}
\begin{figure}[t!]
\includegraphics[width=1.02\columnwidth]{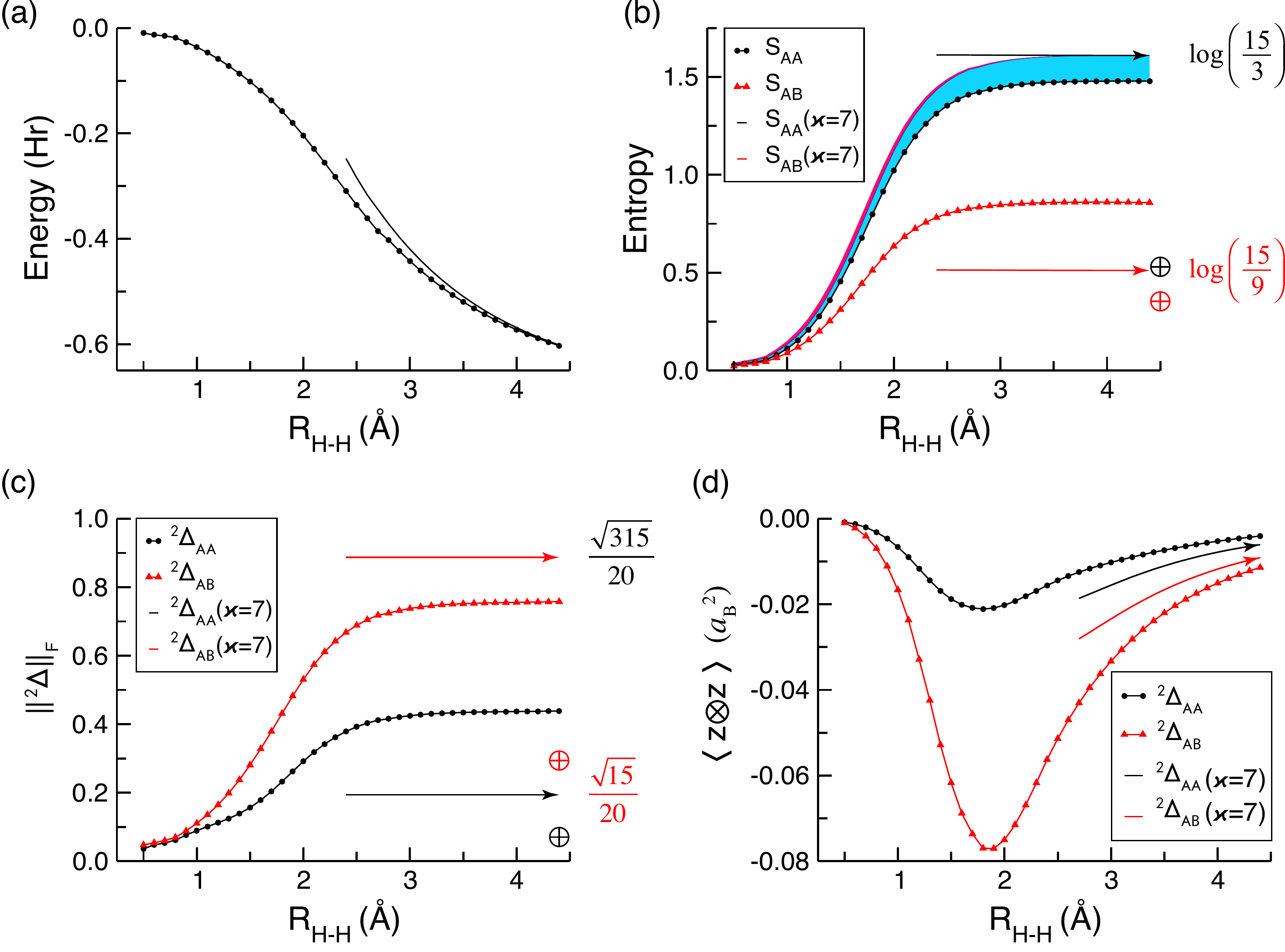}
\caption{(Color online) Full CI calculation for the stretched H$_6$ ring molecule
  performed with aug-cc-pvqz basis set, 30 molecular orbitals included in a correlated
  treatment. The electron correlation increases monotonically as the bonds are
  stretched. This is manifested in the correlational energy (a), the entropies (b), the
  Frobenius norm of the second cumulant matrices (c), and the expectation value of the
  $z\otimes z$ operator (cf. $\langle[\hat S\otimes\hat S]\rangle$ in Eq.~\eqref{eq:res1}
  or $\sum_{ijkl}S_{ij}S_{kl}{}^2\!\Delta^{ik}_{jl}$ in Eq.~\eqref{eq:res2}) (d).
    The entropies are given with respect to the single determinant values $S_{AA}^0$ and
    $S_{AB}^0$. Thin lines with arrows denote results for $\varkappa=7$ state. Notice,
    that this state was only computed for larger interatomic distances where it becomes
    indistinguishable from the ground state (see the energy plot). Blue and red shading
    stands for the first and second Carlen-Lieb bound (Eq.~\eqref{eq:ineq2}). Black and
    red circles denote results for the 3H$^-$ asymptotic state.
\label{fig:H6}}
\end{figure}

We consider two simple multielectron system, the LiH molecule
(Figs.~\ref{fig:LiH_PES},\ref{fig:LiH}) and the H$_6$ ring
(Figs.~\ref{fig:H6_PES},\ref{fig:H6}) and study how different correlation measures vary
when the system deviates from equilibrium geometries. We use our
implementation~\cite{pavlyukh_configuration_2007,pavlyukh_initial_2013} of the algorithm
by Olsen \emph{et al}.~\cite{olsen_determinant_1988} based on the graphical unitary group
approach~\cite{knowles_new_1984} for performing full CI calculations and subsequent
determination of 2-RDMs, coupled cluster and multireference calculations are carried out
with {\sc gaussian 03} program. The systems represent two different scenarios of
electronic correlations adopted in quantum chemistry. The diatomic molecule is a typical
system with importance of \emph{dynamic} correlations. Numerous Slater determinants
contribute to the correlation energy, however, \emph{one} of them is dominant. Thus, even
for stretched geometries the Hartree-Fock (HF) solution represents a valid starting point
for treating electronic correlations with single-reference
methods~\cite{scuseria_assessment_1990}. This is not so for the H$_6$ ring for which
\emph{static} correlations are important. Even for slight deviations from the equilibrium
the wave-function of the system takes a form of a sum of several equally significant
Slater determinants. It was shown by B{\'e}nard and Paldus~\cite{benard_stability_1980}
that restricted HF solutions are unstable with respect to spin-unrestricted perturbations
for a wide range of geometries. This invalidates the restricted Hartree-Fock approach and
also all the single-reference correlated methods based on it. At the same time poses an
interesting question on whether such spin-instabilities could be detected by some
correlation measures.

Let us consider first a simpler case of a \emph{single bond breaking}. It can be easier
treated with quantum chemistry methods as compared, e.~g., with double bond breaking in
O$_2$, C$_2$, or triple bond breaking in N$_2$ dimers. However, the physical mechanisms
are the same. LiH possesses only four electrons (in $1\sigma^22\sigma^2$ configuration of
the ground ${}^1\Sigma^+$ state) which makes it well suited for exact diagonalization
studies. For these rather small full electron calculations we used the 6-311++G(2d,2p)
basis set yielding a total of 37 basis functions.  Even with (6s2p)/[4s2p] contracted
basis functions for the H atom and (12s6p2d)/[5s4p2d] contraction for the Li atom it is
capable of representing a substantial portion of the correlation energy
(Fig.~\ref{fig:LiH_PES}). We used the augmented cc-p5z basis set for the verification of our
results. Stretching the molecule (the equilibrium distance is 1.6~{\AA} based on a CCSD
calculation) leads to the breaking of a single covalent bond and to an increase of the
correlation energy (Fig.~\ref{fig:LiH}).  Our simulations indicate similar behavior for
the entropies and norms of the second cumulant matrices. Notice that the molecule is in
the singlet ground state and therefore there are only two independent spin blocks:
${}^2\Delta_{BB}={}^2\Delta_{AA}$, ${}^2\Delta_{BA}={}^2\Delta_{AB}$. If HCP has the
electric field vector aligned perpendicular to the molecule's axis (in $z$-direction) the
expectation value of the dipole moment vanishes ($\langle z\rangle=0$) and we are in the
situation analyzed above. The correlated part of the survival probability is given by the
expectation value of the $z\otimes z$ two-particle operator and is also computed in $AA$
and $AB$ channels (Fig.~\ref{fig:LiH}). Its dependence on the interatomic distance is
determined by two factors: the norm increase of ${}^2\Delta$ and the matrix elements
reduction of the dipole operator. The latter dominates the behavior at large inter-nuclear
separations.

Let us look now at the manifestly \emph{multi-reference system}, the H$_6$ ring, and
investigate how the correlation measures depend on a single geometric parameter, the
nearest neighbor distance $R_{\mathrm{H-H}}$. Notice, that such symmetric distortion
represents a somewhat artificial situation as the dimerized state possesses a lower ground
state energy. Multireference SCF in the subspace of six electrons and six orbitals is
capable of recovering a major part of the correlation energy and also correctly predicts
the asymptotic state of six unpaired electrons residing on six independent H atoms. In
contrast, RHF method which assumes that each molecular orbital is doubly occupied cannot
produce such asymptotic state and fails shortly above the equilibrium distance. Even more
drastic divergence shows the single reference coupled cluster approach. In fact, it does
not even converge beyond $R_{\mathrm{H-H}}=3.6$~{\AA}. On the same Fig.~\ref{fig:H6_PES}
results of full CI are shown. The potential energy curve runs almost parallel to MCSCF,
however, has a slightly higher energy because not all molecular orbitals (only 30) were
included in the calculation. Both methods nicely converge towards the asymptotic energy of
$-3$~Hartree.  We use the augmented cc-pqz basis in the (7s4p3d2f)/[5s4p3d2f]
contraction. The survival probability was computed for the case of electric field
perpendicular to molecular plane, the expectation value of the dipole moment is zero
($\langle z\rangle=0$). Correlation energy of this system increases steeply as a function
of H-H distance. The norms of the second cumulant matrix and the entropies behave
similarly.  As in the case of LiH, the expectation value of the direct product $z\otimes
z$ is governed by two counteracting factors: increasing ${}^2\Delta$ and reduction of
$\vec z$.  At first sight the spin instability is not manifested clearly in these
correlational measures. They show essentially the same dependence (up to some scaling) for
AA and AB blocks. The difference becomes obvious only by comparing with other electronic
states that have different spin configurations.

In the asymptotic limits some electronic states of the H$_6$ system permit analytical
treatment. Because of the vanishing electron repulsion, 6H is the asymptotic ground state
with four possible values of the total spin.  At large $R_{\mathrm{H-H}}$ the atomic
orbitals of six H atoms are not perturbed by the interaction with other ions or electrons
and each electron is in $1s$ state. Thus, the density matrices can be conveniently
computed in this basis. Since $N_\alpha=N_\beta$ each quantum state of such six electron
system is given by a linear combination of the following Slater determinants
$|\alpha_1\alpha_2\alpha_3\bar\beta_1\bar\beta_2\bar\beta_3\rangle$, where
$\{\alpha_1,\alpha_2,\alpha_3,\beta_1,\beta_2,\beta_3\}$ is a permutation of orbitals 1 to
6 such that $\alpha_1<\alpha_2<\alpha_3$ and $\beta_1<\beta_2<\beta_3$, i.e. $\alpha$ and
$\beta$ string are written in the lexicographic order. The spin part of the basis functions
is indicated by bar for the spin down states. There are in total $\frac{6!}{3!\cdot3!}=20$
such determinants and they can be further divided into four groups with multiplicities
$\varkappa=1,\,3,\,5,\,7$ each containing 5, 9, 5 and 1 state, respectively. All twenty
states are degenerate and we consider a particular example of the $\varkappa=7$ state in
which all the Slater determinants enter with the same coefficient. Other spin multiplicity
states can be easily constructed as explained in Pauncz~\cite{pauncz_spin_1979}. The
two-body density matrix is built from combinatorics factors and evaluates as follows:
\begin{subequations}
\label{eq:2d}
\begin{eqnarray}
^2\!D^{ij}_{ij}&=&
-^2\!D^{ij}_{ji}=\frac{4}{20}(1-\delta_{ij}),\\
^2\!D^{i\bar{j}}_{i\bar{j}}&=&-^2\!D^{i\bar{j}}_{j\bar{i}}=\frac{6}{20}(1-\delta_{ij}),
\end{eqnarray}
\end{subequations}
and vanishes otherwise. Tracing AA or AB blocks leads to the same one-particle density matrix
\begin{eqnarray}
^1\!D^{i}_{j}&=&\frac{1}{N_\alpha-1}\sum_{k=1}^6\,^2\!D^{ik}_{jk}
=\frac{1}{N_\beta}\sum_{k=1}^6\,^2\!D^{i\bar{k}}_{j\bar{k}}=\frac12\delta_{ij}.
\end{eqnarray}
It is clear that $^1\!Q^{j}_{i}=\langle\hat{c}_j\hat{c}^\dagger_i\rangle=\,^1\!D^{i}_{j}$
and $^1\!Q^{i}_{\bar{j}}=\,^1\!D^{i}_{\bar{j}}=0$. One verifies that the expectation value
of the total spin squared operator indeed corresponds to the multiplicity
$\varkappa=2\cdot3+1=7$:
\begin{eqnarray}
\langle \hat{\mathcal{S}}^2\rangle&=&\big\langle
\hat{\mathcal{S}}_{-}\hat{\mathcal{S}}_{+}+\hat{\mathcal{S}}_z^2+\hat{\mathcal{S}}_z\big\rangle
\stackrel{\langle \hat{\mathcal{S}}_z\rangle=0}{=} \Big\langle\sum_i
c^\dagger_{\bar{i}}c_i\sum_j
c^\dagger_{j}c_{\bar{j}}\Big\rangle\nonumber\\ &=&N_\beta-\sum_{i,j=1}^6\,
^2\!D^{j\bar{i}}_{i\bar{j}}=12.
\end{eqnarray}
The cumulant matrix is easily evaluated
\begin{subequations}
\begin{eqnarray}
^2\!\Delta^{ij}_{ij}&=&
-^2\!\Delta^{ij}_{ji}=-\frac{1}{40}(1-\delta_{ij}),\\
^2\!\Delta^{i\bar{j}}_{i\bar{j}}&=&
\frac{1}{40}-\frac{6}{40}\delta_{ij},\\
^2\!\Delta^{i\bar{j}}_{j\bar{i}}&=&
-\frac{6}{40}+\frac{1}{40}\delta_{ij}.
\end{eqnarray}
\end{subequations}
This yields the norms $||^2\!\Delta_{AA}||_F=\frac{\sqrt{15}}{20}$ and
$||^2\!\Delta_{AB}||_F=\frac{\sqrt{315}}{20}$. The expectation values of $z\otimes z$ with
respect to $\tns{2}{\Delta}_{AA}$ and $\tns{2}{\Delta}_{AB}$ are zeros because the dipole
transition moments between $1s$ states vanish due to symmetry as manifested in the
selection rules. Notice, that for finite distances there is a hybridization between these
and higher angular momentum states. Therefore, $z\otimes z$ can be different from zero.

The eigenvalues of the matrices (Eqs.~\eqref{eq:2d}) are
$w^{AA}_{1,\ldots,15}=\frac{1}{15}$, $w^{AA}_{16,\ldots36}=0$, and
$w^{AB}_{1,\ldots,15}=\frac{3}{15}$, $w^{AB}_{16,\ldots36}=0$. After the normalization we
obtain the pair entropies $S_{AA}=S_{AB}=\log{15}$. Similarly, the entropies can be
computed from the 1-RDM, $S_{A}=S_{B}=\log{6}$.

Another interesting case is in which the system dissociates into three hydrogen anions
(H$^-$) and three protons. This is an excited 20-fold degenerate, yet bound state of the
system with the binding energy of just $3\times0.7542$~eV=$3\times0.0277$~Hartree
(measured with respect to the energy of 3 neutral H atoms). The state is extremely
correlated with RHF not being capable of yielding the negative energy (for an overview on
the electronic structure of H$^-$ we refer to A.~R.~P.~Rau~\cite{rau_negative_1996}, in
particular we quote from this work: "The (wave-) function exhibits a radial "in-out"
correlation between the electrons such that when one electron is "in" close to the
nucleus, the other is kept "out"."). We can also view H$^-$ as a partial case of the He
isoelectronic series with the nuclear charge $Z$ being very close to the critical value
$Z_{\mathrm{crit}}\approx0.911$ at which "a quantum phase transition" from a bound to an
unbound two-electron system occurs. A detailed analysis of this system was performed in
the context of strictly correlated-electrons functional~\cite{mirtschink_energy_2014}. The
fact that there are just two electrons in the system allows to exactly compute the
Kohn-Sham (KS) potential by simple inversion of the KS
equation~\cite{umrigar_accurate_1994}. The KS molecular orbitals are eigenfunctions of the
non-interacting Schr{\"o}dinger equation. For the singlet $\varkappa=1$ two-electron state
there is a one-to-one correspondence between the exact density and the Kohn-Sham orbitals
$\psi(\vec r)=\left[\frac{\rho(\vec r)}{2}\right]^{1/2}$. Even though the KS potential can
be explicitly constructed, the density functional theory does not permit to find the
off-diagonal elements of 1-RDM. Therefore, we use again the full CI approach. It is
sufficient to diagonalize the Hamiltonian for a single H$^-$ atom (for such a system the
larger aug-cc-pv6z basis set can be used) and to construct the $\tns{2}{\Delta}$ for the
total system by the observation that it is given by a direct sum of the cumulant matrices
of each subsystem. In order to compute the entropies we reconstruct the 2-RMD from the
cumulant matrix and using the fact that 1-RDM is non-zero only when both indices denote
states of the same H$^-$ anion.

Numerical results for different correlation measures are shown at Fig.~\ref{fig:H6} for
the ground state together with asymptotic values for the $\varkappa=7$ state (obtained
analytically) and for the 3H$^-$ state. The latter is based on the full CI treatment of
H$^-$ and is only valid for $R_{\mathrm{H-H}}\rightarrow\infty$. For finite distances
there is a coupling between the subsystems and one needs to diagonalize the full
Hamiltonian. This is, however, a formidable task since 3H$^-$ is a highly excited state of
the system not amenable to the Lanczos diagonalization.

The blue shaded area on the entropy plot shows the difference between the exact value of
the two-particle entropy and the first Carlen and Lieb estimate,
$2S_1-\ln\left(\frac{2}{1-\mathrm{Tr}\rho_1^2}\right)$. The red shaded area illustrates
the second Carlen and Lieb inequality, $2S_1-\left(\frac{2}{1-e^{-S_1}}\right)$. Since at
the asymptotic limit the 1-RDM and, therefore, $S_1$ and $\mathrm{Tr}\rho_1^2$ are the
same for all twenty degenerate states, same inequalities hold for all of
them. $\varkappa=7$ state saturates the inequalities and hence has the largest possible
two-particle entropy among these states. 3H$^-$, on the other hand, has the smallest
entropies and Frobenius norms out of all states that we considered. For them correlational
measures lie in between these extremes. In fact, all three of them are capable of
discriminating between the spin configurations. The survival probability which is closely
connected to the averaged value of $z\otimes z$ can even be measured experimentally. In
contrast, the 1-RDM and all associated correlational measures are ignorant to the spin
configuration as the example of the asymptotic 6H states shows.
Finally, we note the correlation between the saturation behavior seen in
Fig.~\ref{fig:H6}(b,c) when $R_{H-H}$ is increased beyond $\approx 2.5$~{\AA} and the
flatness at these distances of the corresponding potential energy surface, depicted in
Fig.~\ref{fig:H6_PES}. Even though the electrostatics is less relevant at these distances,
the non-local nature imposed by the (anti)symmetry on the total wave function persists for
the isolated expanding system. This means on the other hand, that the asymptotic values of
the entropy and $||^2\Delta||_F$ delivers short-range (atomistic) information. This
picture might change in a nontrivial way when coupling the electronic and the ionic system
to external baths, in particular if the bath implies some spin-flip scattering.

\section{Conclusions}
We explored theoretically the possibility of accessing the electron correlations using a
pump-probe technique with half-cycle pulses.  The non-resonant kick-type excitation is
appropriate in sofar as it allows for many-particle excitations and can still be amenable
to quasi-analytical treatment. We demonstrated that to the leading order in the field
intensity, the probability for a system to remain and/or recur to the initial state (the
survival probability) can be expressed in terms of 2-RDM and, hence, represents a measure
for electronic correlations. Exact diagonalization studies were performed on two molecules
LiH and H$_6$ -- prototypes of systems with importance from the point view of dynamic and
static correlations. In the dissociative limit the second system permits for an analytical
treatment of some electronic (asymptotic) states for which different correlational
measures and entanglement entropies were computed. Even though these asymptotic subsystems
can be considered as noninteracting, the states are entangled due to the wave-function
anti-symmetry. This is reflected in the two-particle entropy, the Frobenius norm of the
second cumulant matrix $||^2\!\Delta||_F$ and in the expectation value of the $z\otimes z$
operator. The latter can be inferred from the proposed measurement of the survival
probability. Thus corresponding experiments can probe the spin configuration of many-body
systems.
\section{Acknowledgments} The work is supported by DFG-SFB762.
We thank Robert Moshammer for stimulating discussions.
\end{document}